# An algebraic hypothesis about the primeval genetic code

Robersy Sánchez[1, 2] and Ricardo Grau[2]

[1] Research Institute of Tropical Roots, Tuber Crops and Bananas (INIVIT). Biotechnology Group. Santo Domingo. Villa Clara. Cuba.

[2] Center of Studies on Informatics. Central University of Las Villas. Villa Clara. Cuba.

**Classification**:

Applied Mathematics, Evolution.

**Corresponding author:** Robersy  Sánchez

**Email:** robersy@uclv.edu.cu

**Corresponding  address:**  Apartado postal 697. Santa Clara 1. CP 50100. Villa Clara. Cuba.

**Comment:** A part of this article was presented in the international workshop IWOBI'08 at the Universidad Central de Las Villas, Santa Clara, Cuba. February 4-7, 2008.

## Abstract

A plausible architecture of an ancient genetic code is derived from an extended base triplet vector space over the Galois field of the extended base alphabet {D, G, A, U, C}, where the letter D represents one or more hypothetical bases with unspecific pairing. We hypothesized that the high degeneration of a primeval genetic code with five bases and the gradual origin and improvements of a primitive DNA repair system could make possible the transition from the ancient to the modern genetic code. Our results suggest that the Watson-Crick base pairing $G \equiv C$ and $A = U$ and the non-specific base pairing of the hypothetical ancestral base D used to define the sum and product operations are enough features to determine the coding constraints of the primeval and the modern genetic code, as well as, the transition from the former to the later. Geometrical and algebraic properties of this vector space reveal that the present codon assignment of the standard genetic code could be induced from a primeval codon assignment. Besides, the Fourier spectrum of the extended DNA genome sequences derived from the multiple sequence alignment suggests that the called period-3 property of the present coding DNA sequences could also exist in the ancient coding DNA sequences.

# 1. Introduction

Mutations continuously occur in the genomes of all living organisms, at a very low frequency, which tends to be constant for a specific species. Genome instability caused by the great variety of DNA-damaging agents would be an overwhelming problem for cells and organisms if it were not for DNA repair. So, we can ask how the prebiotic "genetic information" without a repair system was conserved. A plausible explanation could be the existence of extra DNA bases with unspecific pairing with the present bases and the possible high degeneration of the primeval genetic code. If we suppose that there was a primeval genetic code then there are not biological reasons to restrict the base alphabet of the primeval RNA and DNA to the present DNA bases. Likewise, there are not biological reasons to restrict the number of codons which code to amino acids. Actually, the environmental conditions of the primeval "cell-like" entities could make possible the relative abundance of different bases and their isomers (1, 2). This richness of bases should increase the success probability of life. The Watson-Crick base pairs $G \equiv C$ and $A = U$ (where each '$-$' symbolizes a hydrogen bond) characterize the present DNA molecule. The accessibility to five or more bases in the "primeval soup" makes plausible the non-Watson-Crick base pairing in a primitive RNA world and later in a primeval DNA molecule (2).

The existence of uncommon bases in the present RNA and DNA molecules could be the relict genetic fingerprint of a molecular evolution process from primordial cells with an extended DNA base alphabet. At present, minor bases of DNA, 5-Methylcytidine occur- in the DNA of animals and higher plants, N-6-methyladenosine in bacterial DNA, and 5-hydroxymethylcytidine in the DNA of bacteria infected with certain bacteriophages. Ribosomal RNAs characteristically contain a number of specially modified nucleotides, including pseudouridine residues, ribothymidylic acid and methylated bases. The abundance of bases with non-specific pairing in a primeval DNA molecule could increase the degeneration of coding apparatus and diminished the error frequency during transcription and translation processes (3).

Orgel (1) summarized the difficulties in prebiotic synthesis of the nucleosides components of RNA (nucleo-base+sugar) and suggested that some of the original bases may not have been purines and pyrimidin. Piccirilli et al. (3) demonstrated that the alphabet can in principle be larger. C. Switzer et al. (4) have shown an enzymatic incorporation of new functionalized bases into RNA and DNA. This expanded the genetic alphabet from 4 to 6 letters, permits new base pairs, and provides RNA molecules with the potential to greatly increase the catalytic power. Actually, a number of alternative base pairs have been proposed. They include isoguanine and isocytosine (3, 5), diaminopurine and U (3), pseudodiaminopyrimidine (ribose bound to C5 rather than N1) and xanthine (3), A and urazole (1,2,4-triazole-3,5-dione) (6, 7). However, according to Levy and Miller, to get greater stability, it may be necessary to use bases other than these pyrimidines (8).

The simpler translation machinery of the mammalian mitochondrial genome suggests that the universal code, as we understand might not have existed at the beginning of the life. Ohno and Epplen (9) proposed that life started with the simpler mitochondria-like code involving fewer species of tRNAs and, therefore, fewer anticodons of greater infidelity with respect to their codon recognition. The last suggestions still presume that the Watson–Crick pairing of A with U and of G with C is retained as the basis of the genetic template recognition and that these bases were readily available on early Earth. Shapiro (10, 11) argued that presumption is not supported by the existing knowledge of the basic chemistry of these substances. Levy and Miller (7) pointed out that the rates of decomposition of the nucleobases A, U, G, C, and T clearly show that these compounds are not stable on a geologic time scale at temperatures much above 0°C (12). Even at 25°C, the hydrolysis rates of the compounds are fast on the geologic time scale. They conclude that unless the origin of life took place extremely rapidly (<100 yr), a high-temperature origin of life may be possible, but it cannot involve adenine, uracil, guanine, or cytosine.

Here, we shall show that the standard genetic code architecture could be derived from an ancient coding apparatus with an extended alphabet of five bases or more bases. Besides, it is shown that all mutational events that take place in the molecular evolution process can be described by means of endomorphisms, automorphisms and translations of a novel DNA sequence vector space over the Galois field *GF* (5).

## 2. Theoretical model

The natural extension of DNA alphabet allows us to define new genetic code algebraic structures similar to those recently published (13-15). In particular, we have defined a new Galois field (GF(5)) over the set of extended RNA alphabet $B$ = {D, A, C, G, U}, where the letter D symbolizes one (or more) alternative hypothetical base(s) or a dummy variable with non-specific pairing present in a primeval RNA and DNA molecules. If a Galois field algebraic structure is defined on the extended base alphabet subject to the constraint A + U = U + A = D and A • U = U • A = G then the sum and product operations can be defined on the set *B*. That is, it is required that bases A and U will be inverses in the sum and product operations with the base G as neutral element for product operation. So, these definitions reflect the Watson-Crick base pairs G ≡ C and A = U distinctive of the present DNA molecule (16) and the non-specific pairing of the ancient hypothetical base(s) D. The definitions of sum and product operations are presented in Table 1. By construction, the Galois field (*B*, +, •) defined in the extended base alphabet is isomorphic to the field of integer's remainder modulo 5

($\mathbb{Z}_5$), a simple representation of GF(5). Explicitly, there is the bijection: D$\leftrightarrow$0, G$\leftrightarrow$1, A$\leftrightarrow$2, U$\leftrightarrow$3, C$\leftrightarrow$4. In order to abbreviate the field $(B, +, \bullet)$ will be also denoted by the symbol $B$.

## 2.1. The extended base-triplet vector space over the Galois field $B$

The extended DNA alphabet naturally leads us to an extended set of base-triplets $XYZ$, i.e. an "extended genetic code". At this point, a new base-triplet vector space over GF(5) can be derived in analogous way to the genetic code vector space over GF(4) as appear in (14, 15). An abelian group on the set of extended triplets set $B^3 = \{XYZ\}$ (see Table 2) can be defined as the direct third power of $(B, +)^3 = (B, +) \times (B, +) \times (B, +)$ of the group $(B, +)$, where the operation "+" is given as appear in Table 1 and $X, Y, Z \in B$. Next, for all elements $\alpha \in$ GF (5) and for all codons $XYZ \in (B^3, +)$, the element

$$\alpha \bullet XYZ = \overbrace{XYZ + XYZ + ... + XYZ}^{\alpha \text{ times}} \in \left( B^3, + \right) \text{ is well defined. As a result, group } (B^3, +) \text{ is}$$

a three-dimensional vector space over GF (5). Likewise, the $N$-dimensional vector space $(B^3)^N$ is obtained.

The bijection D $\leftrightarrow$ 0, G $\leftrightarrow$ 1, A $\leftrightarrow$ 2, U $\leftrightarrow$ 3 and C $\leftrightarrow$ 4 allows us to insert the set $B^3$ into the ordinary three-dimensional vector space $\mathbb{R}^3$ and it can be represented as an ordinary cube or a regular hexahedron with three of its faces contained in the coordinated planes $XY$, $XZ$ and $YZ$ (see Figure 1). The extension of this vector space to the $N$-dimensional vector space $(B^3)^N$ of DNA sequences should lead to new biological-algebraic insights. The extended DNA sequences appear, normally, during the multiple sequence alignment of genes or genomic sequences. The aligned DNA sequences with gaps resulting of the alignment procedure could be analyzed, integrally, as an element of the new vector space.

## 2.2.  Linear transformations in $B^3$ and $(B^3)^N$

Lineal transformations will allow us to study the mutational pathway in the $N$-dimensional space $(B^3)^N$ of DNA sequences. The algebraic operations over the base triplet are equivalent to the derivation of new codons by means of base substitution mutations in the ancestor codon. In particular, we are interested in the automorphisms on $B^3$. Since these transformations are invertible, the mutation reversions are forecasted. In addition, it is well known that the set $G$ of all endomorphisms is a ring ($End(G)$) and the automorphisms subset is a ring $Aut(G) \subset End(G)$. Notice that here the endomorphism $f$ will be an automorphism if, and only if, the determinant of its endomorphism representing matrix ($a_{ij}$) is not equal to the extended base D, i.e. $Det((a_{ii})) \neq$ D. The endomorphism ring $End(G)$ is isomorphic to the ring of all matrices ($a_{ij}$), where $a_{ij} \in GF(5)$ ($i, j = 1,2, 3$), with the traditional matrix operations of sum and product defined over $GF(5)$. Next, for all elements $\alpha \in B$ and for all extended base-triplet $XYZ \in B^3$ and $f \in End(G)$, the element $\alpha \bullet f \in G$ is well defined and the group ($End(G)$, +) will be a vector space over $B$, while the subgroup $Aut(G)$ will be a vector subspace.

Now, we can analyze endomorphisms on the $N$-dimensional vector space $(B^3)^N$ of DNA sequences. The endomorphism ring $End((B^3)^N)$ is, now, isomorphic to the ring of all matrices  ($A_{ij}$), where $A_{ii} \in End(B^3)$ (or $A_{ii} \in Aut(B^3)$) ($i, j = 1, .., N$) and $A_{ij} = 0$ for $i \neq j$, with the traditional matrix operations of sum and product, i.e., the principal diagonal elements are matrices and the non-diagonal elements are null-matrices. Mutations in DNA sequences will correspond to automorphisms when $A_{ii} \in Aut(G_i)$ for all triplets in the DNA sequence. Likewise, the group ($End((B^3)^N)$, +) will be a vector space over $B$.

# 3. Results and discussion

The present genetic code should resemble the ancient one. The reasons reside in the need to minimize the translation and transcription errors, as well as, the effects of induced mutations by the harmful primitive surrounding environments. So, a plausible general vision of the primeval genetic code architecture could look like as cube presented in Fig. 1. This cube encloses the cubic representation of the present genetic code, discussed in (14).

## 3.1. Geometrical model of the primeval genetic code

The geometric features of the cubic representation of the extended triplet set (see Figure 1) suggest the highly degenerated nature of a primeval coding apparatus.

Let D $\notin$ {G, A, U, C} represents the base (or bases) with non-specific pairing with at least two bases of the set {G, A, U, C} in the DNA molecule. The non-specific pairing of base D is reflected in the model, making it the neutral element of $(B, +)$.

In the vector space of the extended triplets, the subset with non-specific pairing $S_{XDZ} = \{XDZ\}$ conforms a two-dimensional vector subspace, which is contained in the $XZ$ plane. Likewise, the bidimensional vector subspaces corresponding to the subsets with non-specific pairing $S_{XYD} = \{XYD\}$ and $S_{DYZ} = \{DYZ\}$ are contained in the coordinated planes $XY$ and $YZ$, respectively (see Figure 1). In particular, the quotient space $B^3/S_{XDZ}$ has the elements:

$$\{ S_{XDZ}, S_{XDZ} + X_G, S_{XDZ} + X_A, S_{XDZ} + X_U, S_{XDZ} + X_C \}$$

Where $X_G$, $X_A$, $X_U$ and $X_C$ are arbitrary elements of the extended triplet subsets: $\{XGY\}$, $\{XAY\}$, $\{XUY\}$ and $\{XCY\}$ respectively. For instance, codons that belong to the subset $\{XUY\}$ may be represented by some of the sums,

$$\{XUY\} = AUC + S_{XDZ} = AUC + \{XDY\} \text{ or } \{XUY\} = GUA + S_{XDZ} = GUA + \{XDY\}$$

which in particular, for codon AUG, means the translations:

$$AUG = AUC + DDA \text{ or } AUG = GUA + GDC$$

where AUC $\in \{XUY\}$, GUA $\in \{XUY\}$, DDA $\in \{XDY\}$ and GDC $\in \{XDY\}$

The quotient space $B^3/S_{XDZ}$ is a partition of the extended triplets set into five equivalence classes or cosets. Every class has the same number of elements $S_{XDZ}$, i.e. 25 extended triplets, which correspond to the five main columns of Table 2. Subsets $S_{XDZ}$ + $X_G = \{XGZ\}$, $S_{XDZ} + X_A = \{XAZ\}$, $S_{XDZ} + X_U = \{XUZ\}$ and $S_{XDZ} + X_C = \{XCZ\}$ are cosets of the vector subspace $S_{XDZ}$ and consequently, they are affine subspaces of the vector space $B^3$, i.e. they are vertical planes with respect to the horizontal plane $XY$ in Figure 1. Thus, codons that code to amino acids with similar properties should belong −in the primeval genetic code− to the same affine subspace and can be obtained from the extended triplets with non-specific pairing $XDZ$ by means of simple translations. If the primordial coding apparatus evolves so as to minimize translation and transcription errors, then the extended triplets of every vertical plane should code, in general, to amino acids with similar physicochemical properties. So, if the mutation process is described by means of translations, then the geometrical-algebraic features of extended triplet set suggest the transition during the molecular evolution process from a high degenerated primeval code to the present code. In particular, the extended triplets with non-specific pairing D$Y$D ($Y \neq$ D) should code to any amino acids encoded by triplets $X_1YX_3$ ($X_1, X_3 \neq$ D) from the same vertical plane; while the extended triplets D$YZ$ ($Y, Z \neq$ D) should code to any amino acid encoded by triplets $XYZ$ ($X \neq$ D). For instance, the extended triplet DAG should encode to any of the amino acids encodes by the codons GAG, AAG, UAG and CAG, which belong to the same horizontal line (see Fig. 1 and Table 2).

While, the extended triplet subsets $\{XDZ\}$ ($X, Z \in B^3$) so as to minimize the mutational effects on protein biological functions could encode for amino acids with middle polarity or for the simplest amino acid glycine. The extended triplet subsets {$XDZ$} could also be a free code subset. The free code regions should decrease the frequency of transcription and translation errors. So, it should not be strange that the primeval genes could have small regions with free codes. Vetsigian et al. (17) advise that the early cell did not require a refined level of tolerance, and so there was no need for a perfect translation. They suggest the concept of a ''statistical protein,'' wherein a given gene can be translated not into a unique protein but instead into a family of related protein sequences. Translation of every primeval mRNA should produce a set of homolog proteins and the most frequent synthesized amino acid sequences should depend on the amino acids cell concentration.

The subset $S_{DDZ}$ = {DDZ} conforms a one-dimensional vector subspace, one of the cube edges, which is inserted in the coordinated $Z$-axis. So, the $S_{DDZ}$ is a vectorial line, which is generated by any extended triplets DDZ with Z ≠ D. While the quotient vector space $B^3/S_{DDZ}$ of the vector space $B^3$ is conformed by extended triplet subsets that have fixed the first and the second nucleotides. As can be noticed in Figure 1, cosets of $S_{DDZ}$ are vertical lines of the cube, which are orthogonal to the plane of extended triplets $XY$D, i.e. to the face inserted in the coordinated plane $XY$. As a result, there are 25 equivalent classes, having every class 5 extended triplets with the first two bases constant. Such arrangement can be observed in Table 2. It can be noticed that, in most of the cases, codons encoding for the same or similar amino acid should belong to the same vertical line. For instance,

AUG + $S_{DDZ}$ = {AUD, AUG, AUA, AUU, AUC}

CAG + $S_{DDZ}$ = {CAD, CAG, CAA, CAU, CAC}

for two of these classes (see Table 2 and Fig 1). Thus, the extended triplets $XY$D should encode to amino acids encoded by codons that belong to its vertical line. This conjecture is based on the non-specific pairing of base D in the third codon position. The first two letters of each codon are the primary determinants of specificity, a feature that has some interesting consequences (16). The primeval life environment could favor the relative abundance of bases different from G, A, U and C. A gradual origin of a primordial coding apparatus without a loss risk of the ancient "genetic information" could be possible by means of the non-specific pairing of these bases. Analogue situation can be found in nature. The anticodons in some tRNAs include the nucleotide inosinate, which contains the uncommon base hypoxanthine. Inosinate can form hydrogen bonds with three different nucleotides U, C, and A. There are 61 different encoding codons. However, organisms do not have 61 tRNA species with all possible anticodons. In 1966, Crick proposed the famous "wobble hypothesis": through non-Watson-Crick base-pairing rules, less tRNA species are needed (16). Based on observations of characterized yeast tRNAs and modified base-pairing rules, Guthrie and Abelson updated and revised the wobble hypothesis and predicted that 46 different tRNA species would be found in yeast, and perhaps in all eukaryotes (18). Now, it is known that most Eukaryotic cells with sequenced genome follow the revised wobble hypothesis almost perfectly (19). In vertebrate mitochondria, however, an unusual set of wobble rules allows the 22 tRNAs to decode all 64 possible codon triplets.

The two-dimensional vector subspaces on $S_{XYD}$ and on $S_{DYZ}$ lead to analogous genetic code partitions. The geometrical arrangement of their extended triplet is also connected with the coding features of standard genetic code and the codon set can be also derived from these subspaces by means of translations. For instance, the horizontal lines {DAG, GAG, AAG, UAG, CAG} and {ADG, AGG, AAG, AUG, ACG} can be obtained from the vectorial lines $S_{XDD}$ and $S_{DYD}$, respectively, by means of the translations:

$$UAG + S_{XDD} = \{DAG, GAG, AAG, UAG, CAG\}$$

$$\text{AUG} + S_{DYD} = \{\text{ADG, AGG, AAG, AUG, ACG}\}$$

In the first case, codons code to similar amino acid which belongs to the same column in Table 2 and, in the second, each codon belongs to a different column of Table 2.

The primeval life environment could favor the relative abundance of bases different from G, A, U and C. A gradual origin of a primordial coding apparatus without a loss risk of the ancient "genetic information" could be possible by means of the non-specific pairing of these bases.

The present architecture of the standard genetic code is not in disagreement with the plausible architecture of an ancient genetic code like to those proposed in Table 2 and in Figure 1. Of course, not necessarily all amino acids should be represented in a primitive coding apparatus but in order to minimize the transcription and translation errors the general coding features, discussed here, must be hold in the genetic code evolution. The gradual origin and improvements of a primitive DNA repair system could make possible the transition from the ancient to the modern genetic code.

## 3.2. Algebraic partitions of the extended triplet set and architecture of the modern genetic codes

Likewise to the three-dimensional vector space over $\mathbb{R}^3$, the elements $X_1Y_1Z_1$, $X_2Y_2Z_2 \in B^3$ are called collinear if $X_2Y_2Z_2 = \lambda \, X_1Y_1Z_1$, where $\lambda \neq D$. The set of all extended triplets can be sorted into 31 subsets of collinear elements (see Table 3). This set has the general form:

$$S_{CT} = \left\{ \left\{ XYZ, X_A Y_A Z_A, X_U Y_U Z_U, X_C Y_C Z_C \right\} \middle| X_\lambda = \lambda X, Y_\lambda = \lambda Y, Z_\lambda = \lambda Z, \lambda \neq D, X \in B, Y \in B, Z \in B \right\}$$

In particular, the subsets of collinear codons have the general form:

$$S_{CC} = \left\{ \left\{ XGZ, X_A AZ_A, X_U UZ_U, X_C CZ_C \right\} \middle| \lambda, X, Z \neq D \right\}$$

The subsets $S_{CC}$ and $S_{CT} \backslash S_{CC}$ comprise subsets with specific and non-specific pairing, respectively (see Table 3). The subset with specific pairing encloses all codons and every collinear subset contains a codon from a different affine subspace. That is, the principal partitions of the standard genetic code $\{\{XGZ\}, \{XAZ\}, \{XUZ\}, \{XCZ\}\}$ are represented in every collinear codon subset, i.e. $XGZ \in \{XGZ\}$, $X_AAZ_A \in \{XAZ\}$, $X_UUZ_U \in \{XUZ\}$ and $X_CCZ_C \in \{XCZ\}$. Moreover, the sum operation defined in Table 1 induces symmetry in every collinear subset (see Table 3),

$$XGZ + X_CCZ_C = DDD \text{ and } X_AAZ_A + X_UUZ_U = DDD$$

We shall say that the extended triplet $X_1Y_1Z_1$ and $X_2Y_2Z_2$ are symmetrical if $X_1Y_1Z_1 + X_2Y_2Z_2 = DDD$. Likewise, in the $N$-dimensional vector space $(B^3)^N$ of DNA aligned sequences with length $N$, we shall say that two DNA sequences $\alpha$, $\beta \in (B^3)^N$ are symmetrical if $\alpha + \beta = \{DDD, DDD,..,DDD\} \in (B^3)^N$.

Next, if we define the product operation between two extended triplet $X_1Y_1Z_1$ and $X_2Y_2Z_2$ as the product by coordinate $X_1Y_1Z_1 \bullet X_2Y_2Z_2 = (X_1 \bullet X_2)(Y_1 \bullet Y_2)(Z_1 \bullet Z_2)$ then the subset of all codons $S_C = \left\{ XYZ \mid X, Y, Z \in \{G, A, U, C\} \right\}$ is closed to product operation, i.e. $(S_C, \bullet)$ is a multiplicative group; while the collinear subset $S_{XXX} = \{GGG, AAA, UUU, CCC\}$ determines the subgroup $(S_{XXX}, \bullet) \subset (S_C, \bullet)$. And the subset of collinear codons is, precisely, the quotient subgroup $S_{CC} = S_C / S_{XXX}$. That is, every collinear codon subset $\{XGZ, X_AAZ_A, X_UUZ_U, X_CCZ_C\} \in S_{CC}$ can be represented as

$$\{XGZ, X_AAZ_A, X_UUZ_U, X_CCZ_C\} = XYZ \bullet S_{XXX}$$

where $XYZ$ is an arbitrary element of $\{XGZ, X_AAZ_A, X_UUZ_U, X_CCZ_C\}$. For instance, the subset of collinear codons $\{AGC, CAU, GUA, UCG\}$ and $\{CGA, UAC, AUG, GCU\}$ can be written, respectively, as

$$AGC \bullet S_{XXX} \text{ and } CGA \bullet S_{XXX}$$

These results suggest that the architecture of the modern genetic codes could be determined from the architecture of an ancient genetic code as those proposed here. Actually, the partition of the modern genetic codes is implicit in the structure $(S_C, \bullet)$. The subset of codons $S_{XGZ} = \{XGZ\}$ determines a subgroup $(S_{XGZ}, \bullet) \subset (S_C, \bullet)$. The quotient subgroup $S_C / S_{XGZ}$ is a partition of the set of codons into 4 equivalence classes:

$$\{S_{XGZ}, XAZ \bullet S_{XGZ}, XUZ \bullet S_{XGZ}, XCZ \bullet S_{XGZ}\}$$

where, $XAZ$, $XUZ$ and $XCZ$ are arbitrary elements of the codon subsets: $\{XAZ\}$, $\{XUZ\}$ and $\{XCZ\}$, respectively, with $X, Z \neq$ D. Every class has the same number of elements found in $S_{XGZ}$, i.e. 16 codons, which are included into the 4 main columns of codons in Table 3. These are the 4 main columns of the modern genetic code tables (see (14)), where each column encloses codons that encodes, in general, to similar amino acids. For instance, codons that belong to the set $\{XUZ\}$ may be represented by some of the products,

$$\{XUZ\} = \text{AUC} \bullet S_{XGZ} \subset \text{AUC} + S_{XDZ}$$

or $\{XUZ\} = \text{GUA} \bullet S_{XGZ} \subset \text{GUA} + S_{XDZ}$

which in particular, for codon AUG, means:

$$\text{AUG} = \text{AUC} \bullet \text{GGC or AUG} = \text{GUA} \bullet \text{AGU}$$

where AUC $\in \{XUZ\}$, GUA $\in \{XUZ\}$, GGC $\in \{XGZ\}$ and AGU $\in \{XGZ\}$

As was pointed out before, codons from the same vertical line code for the same or similar amino acids. Recall that every vertical line encloses five extended triplets; one of them has non-specific pairing. The rest of four codons from every vertical line can be taken as classes from the quotient group $S_C / S_{GGZ}$, where

$$(S_{GGZ}, \bullet) = (\{\text{GGG, GGA, GGU, GGC}\}, \bullet) \subset (S_{XGZ}, \bullet) \subset (S_C, \bullet)$$

As a result, the partition of the modern genetic code tables into 16 subsets of codons that code to the same or similar amino acids is also derived from the ancient architecture proposed here (see (8)). That is, it can be written, for instance,

$$AUG \bullet S_{GGZ} = \{AUG, AUA, AUU, AUC\} \subset AUG + S_{DDZ} = \{AUD, AUG, AUA, AUU, AUC\}$$

$$CAG \bullet S_{GGZ} = \{CAG, CAA, CAU, CAC\} \subset CAG + S_{DDZ} = \{CAD, CAG, CAA, CAU, CAC\}$$

Since, the Watson-Crick base pairing and the non-specific base pairing of the hypothetical ancestral base D are the fundamental features to define the operations of sum and product, our results suggest that these features are enough to determine the coding constraints of the primeval and the modern genetic code, as well as, the transition from the former to the later.

### 3.3. Endomorphisms and automorphisms in $B^3$ and $(B^3)^N$

The substitution mutations between codons can be represented by diagonal automorphisms, i.e. for all codons pairs $X_1Y_1Z_1 \in B^3$ and $X_2Y_2Z_2 \in B^3$ ($X_i$, $Y_i$, $Z_i \in \{$G, A, U, C$\}$) there is an unique automorphism with a representing diagonal matrix ($a_{ij}$) ($a_{ij} \in \{$G, A, U, C$\}$) such that

$$X_2Y_2X_2 = X_1Y_1Z_1(a_{ij}) = X_1Y_1Z_1 \begin{pmatrix} a_{11} & 0 & 0 \\ 0 & a_{22} & 0 \\ 0 & 0 & a_{33} \end{pmatrix}$$

Notice that, there is a bijection between the set of all diagonal automorphism and the set of all codons. Actually, the set of all diagonal automorphism together with the matrix product operation defined on it is an abelian group ($Aut_D$) isomorphic to the group ($S_C$, $\bullet$). As a result, the set of all diagonal automorphisms can be sorted into the same classes of the modern genetic code, as was discussed in the last section.

In general, for all extended base-triplet pairs $X_1Y_1Z_1 \in B^3$ and $X_2Y_2Z_2 \in B^3$ there is, at least, an automorphism with a representing matrix $(a_{ij})$ $(a_{ij} \in B)$ such that

$$X_2Y_2Z_2 = X_1Y_1Z_1(a_{ij}) = X_1Y_1Z_1 \begin{pmatrix} a_{11} & a_{12} & a_{13} \\ a_{21} & a_{22} & a_{23} \\ a_{31} & a_{32} & a_{33} \end{pmatrix}$$

Every endomorphism transforms the elements of one collinear set of extended triplets into another, i.e. if $X_2Y_2Z_2 = \lambda X_1Y_1Z_1$ ($\lambda \neq D$) and $X_3Y_3Z_3 = X_1Y_1Z_1$ $(a_{ij})$ then

$$X_4Y_4Z_4 = X_2Y_2Z_2(a_{ij}) = \lambda X_1Y_1Z_1(a_{ij}) = \lambda X_3Y_3Z_3$$

Now, let $S_k$ be the subset of codons conserving the same base position $k \in \{1, 2, 3\}$ and $Aut(B^3)$ the group of automorphisms. Then, according to the group theory, the set $St(k)$ of automorphisms $f \in Aut(B^3)$ that preserves the base position $k$ is a subgroup of $Aut(B^3)$, i.e.:

$$St(k) = \{f \in G, \text{ such that: } f(X_1Y_2Z_3) \in S_k \text{ and } X_1Y_2Z_3 \in B^3\} \subset Aut(B^3)$$

This subgroup could be called the stabilizer subgroup of the group $Aut(B^3)$ that fixes base position $k$. Next, we take into consideration that most frequent mutations observed in codons preserve the second and the first base position, Accepted mutations on the third base are more frequent than on the first base, and, in turn, these are more frequent than errors on the second base (20-22). These positions, however, are too conservatives with respect to changes in the polarity of the coded amino acids (23). Consequently, the effects of mutations are reduced in the genes and the accepted mutations decreased from the third base to the second. So, we have to expect that most frequent automorphisms observed in the DNA molecules should preserve the second and the first base positions, i.e. most frequent mutations, should belong to the subgroups $St(2)$ and $St(2) \cap St(1)$. In particular, the automorphism subgroup $St(2)$ maps an affine subspace of the quotient subspace $B^3/S_{XDZ}$ into itself, while the subgroup $St(2) \cap St(1)$ maps an affine subspace of the quotient subspace $B^3/S_{DDZ}$ into itself. In particular, the subgroup $Aut_D(2) = Aut_D \cap St(2)$ maps every subset of codons $\{XGY\}$, $\{XAY\}$, $\{XUY\}$ and $\{XCY\}$ of the quotient subgroup $S_C/S_{XGZ}$ into itself, while the subgroup $Aut_D(2) \cap Aut_D(1) = Aut_D \cap (St(2) \cap St(1))$ maps the codon classes of the quotient subgroup $S_C/S_{GGZ}$ into itself.

Hence, the most frequent mutations observed in genes can be described by means of automorphisms that are algebraically connected into group structures over the field $B$.

### 3.4. Some biological remarks

Primordial "cell-like" entities were the "natural lab test" to design the primitive coding sequences starting from large regions of free codes with non-specific pairings. The primordial "cell-like" entities enclosing regions of free codes with non-specific pairings could come into being the progenote −the common ancestral forms of the eukaryotes and the two prokaryotic groups: eubacteria and archaea (25, 26). The newly emerged prebiotic translation machinery had to cope with base sequences that were not preselected to be coding sequences (9).

The regions of free codes with non-specific pairings should confer to the primordial cells high mutability and adaptation capacities. Those ancient cells supporting genome architectures with relatively low frequency of translations and transcription errors could have an advantage over those cells with less fidelity in these molecular processes. The origin and development of the enzymatic mechanisms for recombination and repair could make plausible the transition from the progenote to the primitive eukaryotes and prokaryotes cells. The free coding regions could have been different destinations. In one way, the repair and recombination mechanisms could gradually eliminate the regions of free coding to come into being the primitive prokaryote cells; while, in another way, the DNA replication, repair and recombination mechanisms could gradually replace the free coding regions with bases G, A, T and C to come into being new coding regions or the primitive introns found in eukaryotes cells. Szathmáry pointed out that copying fidelity and metabolic efficiency change with the size of the genetic alphabet (24). Following Szathmáry idea, the emergent enzymatic mechanism for DNA replication, repair and recombinantion in the primeval cells could favor those ancient cells with small alphabet size to come into being the present alphabet with four bases.

### 3.5. The power spectra of extended DNA genomic sequences

It has been pointed out that the relative height of the peak at frequency $f = 1/3$ in the Fourier spectrum is a good discriminator of coding potential [27-28]. This feature is the called period-3 property of a DNA sequence and it has been used to detect probable coding regions in DNA sequences. Here, we show evidences that the called period-3 property of DNA sequences it is also present in the aligned coding regions of DNA genome sequences including the gaps, which are replaced by the hypothetical base D.

The complex representation $C_5 = \left\{ Exp\left( -\dfrac{2\pi i x}{5} \right) \Big| x \in Z \right\}$ of $\mathbb{Z}_5$ is also isomorphic group to

(B, +). Thus, we can represent the bases of the DNA aligned sequences by the elements of $C_5$, i.e. there is the bijection: $\mathbf{D} \leftrightarrow 1$,

$$G \leftrightarrow Exp\left( -\frac{2\pi i}{5} \right), A \leftrightarrow Exp\left( -\frac{4\pi i}{5} \right), U \leftrightarrow Exp\left( -\frac{6\pi i}{5} \right) \text{ and } C \leftrightarrow Exp\left( -\frac{8\pi i}{5} \right)$$

The discrete Fourier transform of a complex representation of a DNA genomic sequence of length $N$ is defined to be

$$F\left( s/N \right) = \frac{1}{\sqrt{N}} \sum_{n=0}^{N-1} g_n e^{\frac{2\pi i}{N} s n} \qquad (1)$$

where $s=0,\ldots,N\text{-}1$ and the frequency $f = s/N$. The power spectrum is given by $S(f) = |F(s/N)|^2$.

The period-3 property was found in HIV-1 whole genomics sequences. The power spectra of the three aligned HIV-1 genomes are presented in Fig 2. The three aligned sequences keep the coding pattern with a peak at frequency 2/3. Thus, despite to the non-protein-coding regions (at the beginning and at the end of these sequences) and the superposition of genes, the period-3 property can be detected.

## 4. Conclusions

The present genetic code architecture could be derived from an ancient coding apparatus with an extended alphabet of five bases. The Watson-Crick base pairing and the non-specific base pairing of the hypothetical ancestral base D used to define the sum and product operations are enough features to determine the coding constraints of the primeval and the modern genetic code, as well as, the transition from the former to the later.

The necessary minimization of translation and transcription errors, as well as, the minimization of the effects of induced mutations by the harmful primitive surrounding environments must imply the resemblance of the ancient and the present genetic codes. So, the transition from an ancient DNA coding sequence to the present could be biologically plausible and mathematically determined by an ancient coding apparatus highly degenerated.
Besides, the Fourier spectrum of the extended DNA genome sequences derived from the multiple sequence alignment suggests that the called period-3 property of the present coding DNA sequences could also exist in the ancient coding DNA sequences.

# Tables and Figures

**Table 1**. Operation tables of the Galois field on the ordered set of the extended bases alphabet *B*={D, G, A, U, C}.

| Sum | | | | | | Product | | | | | |
|---|---|---|---|---|---|---|---|---|---|---|---|
| + | D | G | A | U | C | • | D | G | A | U | C |
| D | D | G | A | U | C | D | D | D | D | D | D |
| G | G | A | U | C | D | G | D | G | A | U | C |
| A | A | U | C | D | G | A | D | A | C | G | U |
| U | U | C | D | G | A | U | D | U | G | C | A |
| C | C | D | G | A | U | C | D | C | U | A | G |

**Table 2**. Extended base-triplet set

| | No | D | No | G | aa¹ | No | A | aa | No | U | aa | No | C | aa | |
|---|---|---|---|---|---|---|---|---|---|---|---|---|---|---|---|
| D | 0 | DDD | 25 | DGD | | 50 | DAD | | 75 | DUD | | 100 | DCD | | D |
| | 1 | DDG | 26 | DGG | | 51 | DAG | | 76 | DUG | | 101 | DCG | | G |
| | 2 | DDA | 27 | DGA | | 52 | DAA | | 77 | DUA | | 102 | DCA | | A |
| | 3 | DDU | 28 | DGU | | 53 | DAU | | 78 | DUU | | 103 | DCU | | U |
| | 4 | DDC | 29 | DGC | | 54 | DAC | | 79 | DUC | | 104 | DCC | | C |
| G | 5 | GDD | 30 | GGD | | 55 | GAD | | 80 | GUD | | 105 | GCD | | D |
| | 6 | GDG | 31 | GGG | G | 56 | GAG | E | 81 | GUG | V | 106 | GCG | A | G |
| | 7 | GDA | 32 | GGA | G | 57 | GAA | E | 82 | GUA | V | 107 | GCA | A | A |
| | 8 | GDU | 33 | GGU | G | 58 | GAU | D | 83 | GUU | V | 108 | GCU | A | U |
| | 9 | GDC | 34 | GGC | G | 59 | GAC | D | 84 | GUC | V | 109 | GCC | A | C |
| A | 10 | ADD | 35 | AGD | | 60 | AAD | | 85 | AUD | | 110 | ACD | | D |
| | 11 | ADG | 36 | AGG | R | 61 | AAG | K | 86 | AUG | M | 111 | ACG | T | G |
| | 12 | ADA | 37 | AGA | R | 62 | AAA | K | 87 | AUA | I | 112 | ACA | T | A |
| | 13 | ADU | 38 | AGU | S | 63 | AAU | N | 88 | AUU | I | 113 | ACU | T | U |
| | 14 | ADC | 39 | AGC | S | 64 | AAC | N | 89 | AUC | I | 114 | ACC | T | C |
| U | 15 | UDD | 40 | UGD | | 65 | UAD | | 90 | UUD | | 115 | UCD | | D |
| | 16 | UDG | 41 | UGG | W | 66 | UAG | - | 91 | UUG | L | 116 | UCG | S | G |
| | 17 | UDA | 42 | UGA | - | 67 | UAA | - | 92 | UUA | L | 117 | UCA | S | A |
| | 18 | UDU | 43 | UGU | C | 68 | UAU | Y | 93 | UUU | F | 118 | UCU | S | U |
| | 19 | UDC | 44 | UGC | C | 69 | UAC | Y | 94 | UUC | F | 119 | UCC | S | C |
| C | 20 | CDD | 45 | CGD | | 70 | CAD | | 95 | CUD | | 120 | CCD | | D |
| | 21 | CDG | 46 | CGG | R | 71 | CAG | Q | 96 | CUG | L | 121 | CCG | P | G |
| | 22 | CDA | 47 | CGA | R | 72 | CAA | Q | 97 | CUA | L | 122 | CCA | P | A |
| | 23 | CDU | 48 | CGU | R | 73 | CAU | H | 98 | CUU | L | 123 | CCU | P | U |
| | 24 | CDC | 49 | CGC | R | 74 | CAC | H | 99 | CUC | L | 124 | CCC | P | C |

¹The one-letter symbol of amino acids.

**Table 3**. Subsets of collinear extended base triplets. The subsets have been sorted into subsets of extended triplets with specific and non-specific pairing. In each subset, four extended triplets are found.

| Collinear extended triplets | | | | | | | |
|---|---|---|---|---|---|---|---|
| Specific pairing (codons) | | | | Non-specific pairing | | | |
| GGG | AAA | UUU | CCC | DDG | DDA | DDU | DDC |
| GGA | AAC | UUG | CCU | GDD | ADD | UDD | CDD |
| GGU | AAG | UUC | CCA | GDG | ADA | UDU | CDC |
| GGC | AAU | UUA | CCG | GDA | ADC | UDG | CDU |
| AGG | CAA | GUU | UCC | GDU | ADG | UDC | CDA |
| AGA | CAC | GUG | UCU | GDC | ADU | UDA | CDG |
| AGU | CAG | GUC | UCA | DGD | DAD | DUD | DCD |
| AGC | CAU | GUA | UCG | DGG | DAA | DUU | DCC |
| UGG | GAA | CUU | ACC | DGA | DAC | DUG | DCU |
| UGA | GAC | CUG | ACU | DGU | DAG | DUC | DCA |
| UGU | GAG | CUC | ACA | DGC | DAU | DUA | DCG |
| UGC | GAU | CUA | ACG | GGD | AAD | UUD | CCD |
| CGG | UAA | AUU | GCC | GAD | ACD | UGD | CUD |
| CGA | UAC | AUG | GCU | GUD | AGD | UCD | CAD |
| CGU | UAG | AUC | GCA | GCD | AUD | UAD | CGD |
| CGC | UAU | AUA | GCG | | | | |

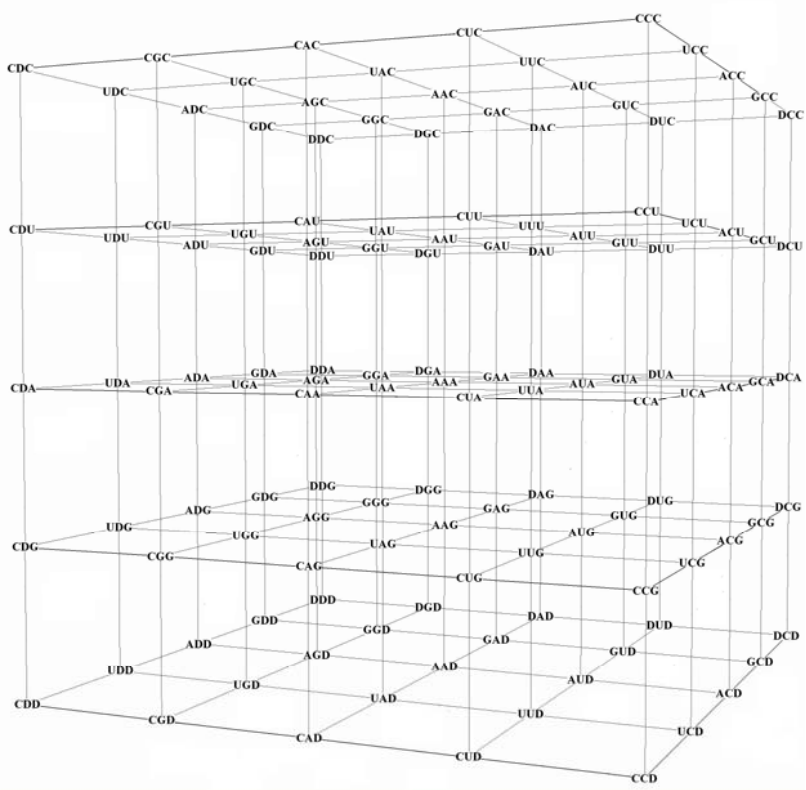

**Fig. 1**. Cubic representation of the extended genetic code.

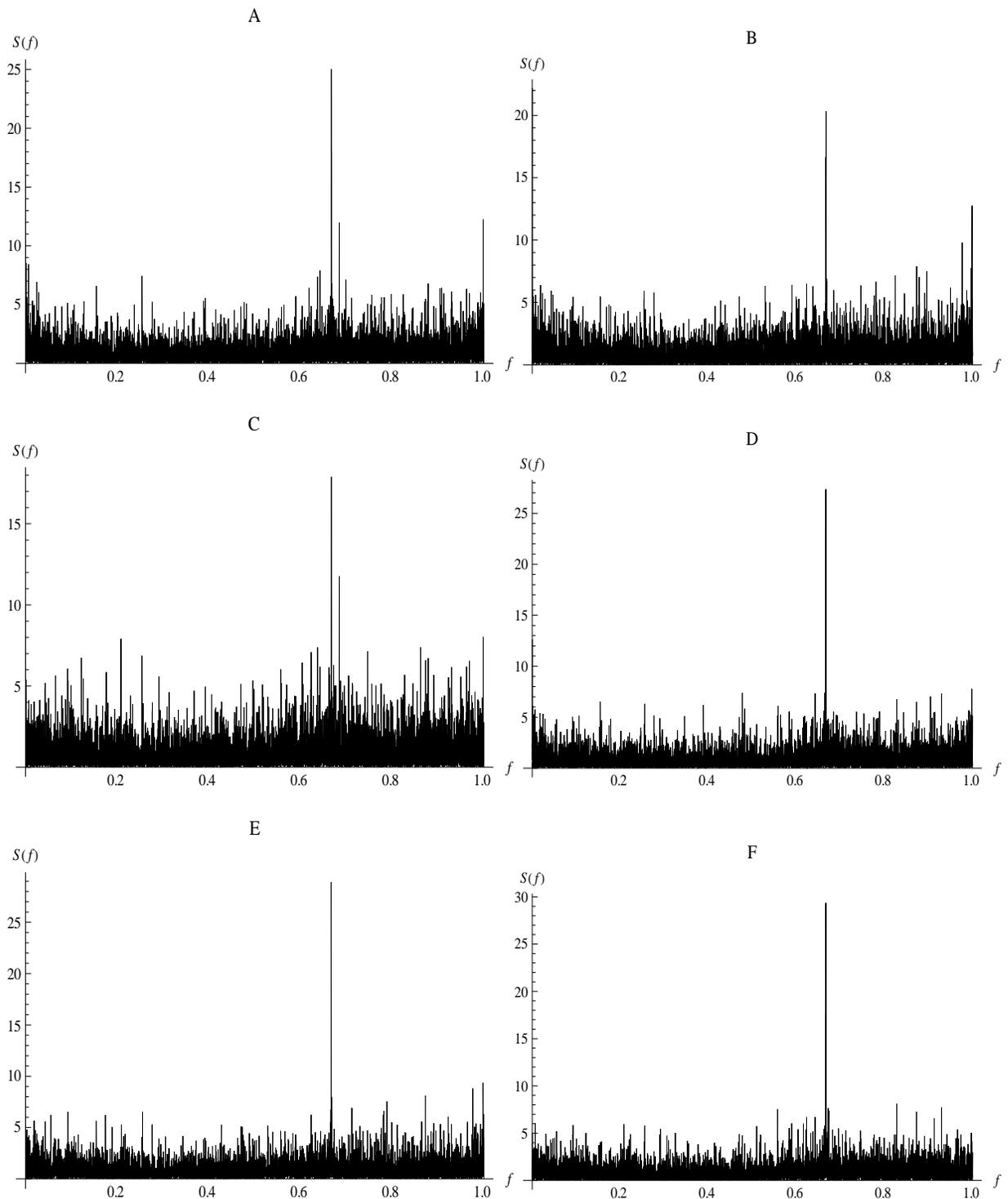

**Fig.** 2. Power spectra of three aligned HIV-1 whole genomes. A, B and C: the power spectra corresponding to the HIV-1 whole genomes sequences with GenBank accession numbers K03455.1 (HXB2), AB023804.1 and U51188.1, respectively. D, E and F: the power spectra corresponding to the same HIV-1 genomes taken from the base position 700 to 9580.